
\magnification=\magstep1

\font\bfw=cmbx12
\hoffset=.4truecm
\voffset= 1truecm
\hsize=15truecm
\vsize=24truecm
\leftskip=0.9cm
\baselineskip=12pt
\hfuzz=18pt
\parindent=1truecm
\parskip=0.2truecm
\parskip=0.2truecm
\def\sqr#1#2{{\vcenter{\vbox{\hrule height.#2pt
        \hbox{\vrule width.#2pt height#1pt \kern#1pt
           \vrule width.#2pt}
        \hrule height.#2pt}}}}

\baselineskip=14pt
\line{\hfill                                                    HEP-TH/9411106}
\line{\hfill                                                 LPTHE-ORSAY 94/90}
\line{\hfill                                                         A337.1194}
\line{\hfill                                                  ENSLAPP-A-496/94}
\vskip 4truecm
\centerline{\bfw  REPRESENTATIONS OF SO(5)$_{\displaystyle{q}}$
AND NON-MINIMAL}
\centerline{\bfw  q-DEFORMATION}
\smallskip
\centerline{B. Abdesselam $^{a,c,}$\footnote{$^{1}$}{abdusa@qcd.th.u-psud.fr},
 D. Arnaudon $^{b,}$\footnote{$^{2}$}{arnaudon@lapphp1.in2p3.fr},
A. Chakrabarti$^{c,}$\footnote{$^{3}$}{chakra@orphee.polytechnique.fr}}
\vskip 1truecm
\centerline{\it $^{a}$ Laboratoire
de Physique Th{\'e}orique et Hautes {\'E}nergies}
\centerline{\it Batiment 211,
91405 Orsay Cedex, France.}
\smallskip
\centerline{\it $^{b}$
ENSLAPP \footnote{$^{4}$}{URA 14-36 du CNRS, associ\'ee \`a l'E.N.S. de Lyon,
et au L.A.P.P. d'Annecy-le-Vieux.},
Chemin de Bellevue BP 110,}
\centerline{74941 Annecy-le-Vieux Cedex, France.}
\smallskip
\centerline{\it $^{c}$
Centre de Physique Th{\'e}orique,
Ecole Polytechnique}
\centerline{\it 91128 Palaiseau Cedex, France.}
\centerline{\it Laboratoire Propre du CNRS UPR A.0014}%
\vskip 1truecm


\noindent
{\bf Abstract}

Representations of $SO(5)_{q}$ can be constructed on bases such that either
the Chevalley triplet $(e_{1},\;f_{1},\;h_{1})$ or $(e_{2},\;f_{2},\;h_{2})$
has the standard $SU(2)_{q}$ matrix elements. The other triplet in each cases
has a more complicated action. The $q$-deformation of such representations
present striking differences. In one case a {\bf non-minimal} deformation is
found to be essential. This is explained and illustrated below. Broader
interests of a parallel use of the two bases are pointed out.

\eject

The $q$-deformation of representations of non-simply laced Lie algebras (with
roots of unequal length ) present special problems. This is illustrated by
comparing, for particular cases, the respective $q$-deformations of
irreducible representations of $SO(5)$ in two bases. Imposing the standard
$SU(2)$ representations for the triplets of Chevalley generators associated
to the shorter and the longer root of $SO(5)$ by turn lead to surprisingly
different consequences concerning $q$-deformation. Irreducible
representations of $SO(5)$ are characterized by two invariant parameters
$n_{1}$ and $n_{2}$ ($n_{1}\geq\;n_{2}$, both integer or half-integer). In
this note we will consider only the cases
$$
\openup 0.5mm
\leqalignno{
& n_{2}=\;0,\;\;\;\;\hbox{and} &\;\;\;\;\;\;\;\;\;\;\;\;\;\;\;\;\;\;\;\;\;
\;\;\;\;\;\;\;{\bf(1)}\cr
& n_{2}=n_{1} &\;\;\;\;\;\;\;\;\;\;\;\;\;\;\;\;\;\;\;\;\;
\;\;\;\;\;\;\;{\bf(2)} \cr}
$$

Upto now only for these two cases the solutions are complete. But even within
such restrictions remarkable features arise. For $n_{2}=n_{1}$ one encounters
an example (defined below) of {\bf non-minimal} $q$-deformation which is our
main result here. The case $n_{2}=0$, needing essentially {\bf minimal}
deformation serves as a contrast. By {\bf minimal} $q$-deformation we
mean [1] introduction of $q$-brackets for each factor in the classical matrix
elements of the Chevalley generators acting on a suitably parametrized set
of basis states. {\bf Non-minimal} means a departure from this involving
subtler, more complicated $q$-deformations of some factors giving back again,
of course, the same classical limit. The significances of these definitions
will be more explicit after the examples to follow.

The Chevalley generators consist of two triplets $(e_{1},\;f_{1},\;h_{1})$,
$(e_{2},\;f_{2},\;h_{2})$ corresponding to the roots 1 and 2 respectively.
The standard Drinfeld-Jimbo construction for $SO(5)_{q}$ is, with commuting
Cartan generators $q^{\pm h_{1}}$, $q^{\pm h_{2}}$,
$$
\openup 2mm
\eqalign{
&q^{\pm h_{1}} e_{1} = q^{\pm 1} e_{1} q^{\pm h_{1}} ,
\;\;q^{\pm h_{1}} f_{1} = q^{\mp 1} f_{1} q^{\pm h_{1}} \cr
&q^{\pm 2 h_{2}} e_{1} = q^{\mp 1} e_{1} q^{\pm 2h_{2}} ,
\;\;q^{\pm 2 h_{2}} f_{1} = q^{\pm 1} f_{1} q^{\pm 2h_{2}} \cr
& q^{\pm h_{1}} e_{2} = q^{\mp 1} e_{2} q^{\pm h_{1}} ,
\;\;q^{\pm h_{1}} f_{2} = q^{\pm 1} f_{2} q^{\pm h_{1}} \cr
& q^{\pm h_{2}} e_{2} = q^{\pm 1} e_{2} q^{\pm h_{2}} ,
\;\;q^{\pm h_{2}} f_{2} = q^{\mp 1} f_{2} q^{\pm h_{2}} \cr
& [e_{1} , f_{2}] = 0,\;\;[e_{2} , f_{1}] = 0 \cr
&[ e_{1},f_{1} ]=\;[2 h_{1}]\equiv
\left ({q^{2 h_{1}}-q^{-2 h_{1}}\over  q - q^{-1}} \right) \cr}
$$
$$
\openup 2mm
\eqalign{
&[ e_{2},f_{2} ]=\;[2 h_{2}]_{2}\equiv
\left ({q^{4 h_{2}}-q^{-4 h_{2}}\over  q^{2} - q^{-2}} \right) \cr
& e_{2} e_{3}^{(\pm)} = q^{\pm 2} e_{3}^{(\pm)} e_{2} \cr
& f_{3}^{(\pm)}  f_{2} = q^{\pm 2} f_{2} f_{3}^{(\pm)} \cr
& [ e_{1}, e_{4}]=0,\;\;[f_{1},f_{4}]=0 \cr} \eqno(1)
$$
where we have defined
$$
\openup 2mm
\eqalign{
& e_{3}^{(\pm)} = q^{\pm 1} e_{1} e_{2}- q^{\mp 1} e_{2} e_{1}, \cr
& f_{3}^{(\pm)} = q^{\pm 1} f_{2} f_{1}- q^{\mp 1} f_{1} f_{2}, \cr
& e_{4}= q^{-1} e_{1} e_{3}^{(+)} - q\;e_{3}^{(+)} e_{1}=
q\;e_{1} e_{3}^{(-)} - q^{-1} e_{3}^{(-)} e_{1}, \cr
& f_{4}= q^{-1} f_{3}^{(+)} f_{1} - q\;f_{1} f_{3}^{(+)}=
q\;f_{3}^{(-)} f_{1} - q^{-1} f_{1} f_{3}^{(-)}. \cr} \eqno(2)
$$
The coproducts, counits and antipodes are the standard ones.

For subsequent, convenient, use we define also
$$
\openup 2mm
\eqalign{
&q^{\pm M} = q^{\pm h_{1}},\;\;\;q^{\pm (K-M)} = q^{\pm 2h_{2}} \cr
&q^{\pm M_{2}} = q^{\pm {1\over 2}(K-M)}=q^{\pm h_{2}} \cr
&q^{\pm M_{4}} = q^{\pm {1\over 2}(K+M)}=q^{\pm (h_{1}+h_{2})} \cr} \eqno(3)
$$

The second order Casimir operator is $[1]$
$$
\openup 2mm
\eqalign{
A &= {1\over [2]} \Bigl\lbrace \bigl( f_{1}e_{1}+
[M][M+1]\bigl) {[2 K+3]_{2}\over [2 K+3]} + [K][K+3]\Bigr\rbrace \cr
&+\bigl(f_{2}e_{2}+{1\over [2]^{2}} f_{4}e_{4}\bigl)+
{1 \over [2]^{2}}\bigl(f_{3}^{(+)}e_{3}^{(+)}
q^{2M+1} + f_{3}^{(-)}e_{3}^{(-)}q^{-2M-1}\bigl). \cr} \eqno(4)
$$

Though we will need in the following only the restricted cases mentioned
before we state here the general result that on the space of states spanning
the irreducible representation ($n_{1},\;n_{2}$)
$$
\eqalign{
A & = {1\over [2]} \Bigl\lbrace [n_{1}][n_{1}+3]+
[n_{2}][n_{2}+1]{[2n_{1}+3]_{2}\over [2n_{1}+3]}\Bigl\rbrace\;
\hbox{\bfw{\bfw{1}}}. \cr} \eqno(5)
$$
where \hbox{\bfw{\bfw{1}}} is the identity. (For $n_{2}=0$, ${1\over 2}$,
$n_{1}$ this reduces to the results in[1].)

Our aim is to compare two bases for irreducible representations
($n_{1},\;n_{2}$) defined as follows.

{\bf BASIS (1):} let
$$
\openup 2mm
\eqalign{
&q^{\pm M } \vert j\;m\;k\;l>
= q^{\pm m } \vert j\;m\;k\;l> \cr
&q^{\pm K } \vert j\;m\;k\;l>
= q^{\pm k} \vert j\;m\;k\;l> \cr
&e_{1} \vert j\;m\;k\;l>
= ([j-m]\;[j+m+1])^{1/2}\vert j\;m+1\;k\;l > \cr
&e_{2} \vert j\;m\;k\;l>= \cr
&([j-m+1][j-m+2])^{1/2}\sum_{\scriptstyle l'}\;a(j,k,l,l')
\vert j+1\;m-1\;k+1\;l'> \cr
&+ (\;[j+m][j+m-1])^{1/2}\sum_{\scriptstyle l'}\;b(j,k,l,l')
\vert j-1\;m-1\;k+1\;l'> \cr
&+(\;[j+m][j-m+1])^{1/2}\sum_{\scriptstyle l'}\;c(j,k,l,l')
\vert j\;m-1\;k+1\;l'> \cr} \eqno(6)
$$

We will consider (for generic $q$) only {\bf real} matrix elements, when for
any two states $\vert x>,\;\vert y>$,
$$
\eqalign{
<x\vert f_{i} \vert y> = <y\vert e_{i} \vert x>
\;\;\;\;\;\;\;\;\;\;\;\;\;( i= 1,\;2) \cr}
$$

The domains of the indices have been obtained. The patterns of multiplicities
are subtle. They are presented below without the derivations.

(i) For ($n_{1},\;n_{2}$) integers
$$
\openup 2mm
\eqalign{
&j = 0,\;1,\cdots,\;n_{1}-1,\;n_{1} \cr
&m = -j,\;-j+1, \cdots,\;j-1,\;j \cr
&k = -l,\;-l+2, \cdots,\;l-2,\;l \cr
&l = 0,\;1,\;2\;\cdots \cr
&j+l = n_{1}-n_{2},\;n_{1}-n_{2}+1, \cdots,\;n_{1}+n_{2} \cr
&j-l-{1\over2}\bigl(1-(-1)^{n_{1}+n_{2}-j-l}\bigl)=-n_{1}+n_{2},
\;-n_{1}+n_{2}+2, \cdots,\;n_{1}-n_{2} \cr} \eqno(7)
$$
(for the comparing with (2.14) of $[1]$ note that when $n_{1}=n_{2}$,
$l=j,\;j-1$ for $j>0$ and $l=0$ for $j=0$.)

(ii) For ($n_{1},\;n_{2}$) half integers
$$
\openup 2mm
\eqalign{
&j = {1\over 2},\;{3\over 2},\cdots,\;n_{1}-1,\;n_{1} \cr
&m = -j,\;-j+1, \cdots,\;j-1,\;j \cr
&k = -l,\;-l+1, \cdots,\;l-1,\;l \cr
&l = {1\over 2},\;{3\over 2},\cdots \cr
&j+l = n_{1}-n_{2}+1,\;n_{1}-n_{2}+3, \cdots,\;n_{1}+n_{2} \cr
&j-l=-n_{1}+n_{2},\;-n_{1}+n_{2}+2, \cdots,\;n_{1}-n_{2}. \cr} \eqno(8)
$$

Upto now the solutions for the reduced matrix elements $a$, $b$, $c$
satisfying all the necessary algebraic constraints have been obtained $[1]$
for the cases
$$
n_{2}=0,\;{1\over2},\;n_{1}
$$
when there is no multiplicity due to $l$ and one can consider states labelled
$\vert j\;m\;k>$. For comparison with the case to follow we reproduce here,
briefly, the results for $n_{2}=0$ and $n_{1}=n_{2}$ (for $n_{2}={1\over 2}$,
see $[1]$).

To {\bf start with consider only generic $q$} (real, positive). For $n_{2}=0$,
$n_{1}=n$;
$$
\openup 2mm
\eqalign{
&a(j,k) = (q + q^{-1})^{-1} \Biggl({[n-j-k]\;[n+j+k+3] \over [2j +
1]\;[2j + 3]}\Biggr)^{1/2} \cr
&b(j,k) = (q + q^{-1})^{-1} \Biggl({[n + j - k + 1]\;[n - j + k + 2] \over
[2j - 1]\;[2j + 1]}\Biggr)^{1/2} \cr
&c(j,k) = 0 \cr
}\eqno (9)
$$
where
$$
\openup 2mm
\eqalign{
&j = 0, 1, 2, \cdots, n \cr
&k = n-j, n-j-2, \cdots, -(n-j-2), -(n-j) \cr
&m = j,j-1,\cdots, -(j-1),-j. \cr}
$$

For $n_{2}=n_{1}=n$ (integer or half integer)
$$
\openup 2mm
\eqalign{
&a(j,k) =(q+q^{-1})^{-1} \Biggl({[n-j]_{2}\;[n+j+2]_{2}\;[j+k+1]\;[j+k+2]
\over [2j+3]\;[2j+1]\;[j+1]_{2}^{2}}\Biggl)^{1/2} \cr
&b(j,k) =(q+q^{-1})^{-1} \Biggl({[n-j+1]_{2}\;[n+j+1]_{2}\;[j-k]\;[j-k-1]
\over [2j+1]\;[2j-1]\;[j]_{2}^{2}}\Biggl)^{1/2} \cr
&c(j,k) =(q+q^{-1})^{-1} [n+1]_{2} {([j-k]\;[j+k+1])^{1/2} \over [j+1]_{2}
\;[j]_{2}} \cr} \eqno(10)
$$
where
$$
\openup 2mm
\eqalign{
&j = n,n-1,\cdots,0(1/2) \cr
&k = j,j-1,\cdots,-(j-1),-j \cr
&m = j,j-1,\cdots,-(j-1),-j. \cr}
$$

Apart from the limiting values of $n_{2}$ mentioned (the lowest $0$ and
${1\over 2}$ and the highest $n_{1}$) not even the {\bf classical}
representations have yet been obtained for this basis. (see the detailed
discussion and comparison of the situation with that in the Gelfand-Zetlin
basis [2] given in [1].) But for the $n_{2}$ values mentioned above setting
$q=1$ and comparing with the generic $q$-case one sees essentially an example
of {\bf minimal} $q$-deformation. The only effect of unequal roots is the
appearence of $[x]_{2}$ brackets along with $[x]$'s.

{\bf BASIS (2):} Consider now the following basis states ($\varepsilon=\pm 1,\;
\varepsilon'=\pm 1$)
$$
\openup 3mm
\eqalign{
&q^{\pm M_{2}} \vert j_{2}\;m_{2}\;j_{4}\;m_{4}>
= q^{\pm m_{2}} \vert j_{2}\;m_{2}\;j_{4}\;m_{4}> \cr
&q^{\pm M_{4}} \vert j_{2}\;m_{2}\;j_{4}\;m_{4}>
= q^{\pm m_{4}} \vert j_{2}\;m_{2}\;j_{4}\;m_{4}> \cr
&e_{2} \vert j_{2}\;m_{2}\;j_{4}\;m_{4} >
= ([j_{2} - m_{2}]_{2} [j_{2}+m_{2}+1]_{2})^{1/2}
\vert j_{2}\;m_{2}+1\;j_{4}\;m_{4} > \cr
&e_{1} \vert j_{2}\;m_{2}\;j_{4}\;m_{4}>=
\sum_{\scriptstyle \epsilon,\; \epsilon '}\;\;
([j_{2}-\epsilon\;m_{2}+{1+\epsilon \over 2}]_{2})^{1/2}\;\times \cr
&c_{(\epsilon,\epsilon ')}(j_{2},j_{4},m_{4})
\;\vert j_{2}+{\epsilon \over 2}\;\;m_{2}-{1\over 2}\;\;j_{4}+
{\epsilon ' \over 2}\;\;m_{4}+{1\over 2}> \cr} \eqno(11)
$$
with, again,
$$
\eqalign{
<x\vert f_{i} \vert y> = <y\vert e_{i} \vert x>
\;\;\;\;\;\;\;\;\;\;\;\;\;( i= 1,\;2) \cr}
$$

The domain of the indices (again for generic $q$ ) are
$$
\openup 2mm
\eqalign{
&j_{2} = 0,\;{1\over 2},\;1,\cdots,\;{n_{1}+n_{2}\over 2} \cr
&j_{4} = 0,\;{1\over 2},\;1,\cdots,\;{n_{1}+n_{2}\over 2} \cr
&m_{2} = -j_{2},\;-j_{2}+1, \cdots,\;j_{2}-1,\;j_{2} \cr
&m_{4} = -j_{4},\;-j_{4}+1, \cdots,\;j_{4}-1,\;j_{4} \cr}
$$
such that
$$
\openup 2mm
\eqalign{
&j_{2}+j_{4} = n_{2},\;n_{2}+1,\cdots,\;n_{1}\cr
&j_{2}-j_{4} = -n_{2},\;-n_{2}+1,\cdots,\;n_{2} \cr} \eqno(12)
$$

Now for $q=1$ a complete solution for the reduced elements
$c_{\epsilon,\epsilon '}$ is available. This is the representation of
Hughes $[3]$. Though it does not seem to be explicitly noted in the paper,
the {\bf shift operators} of $[3]$ correspond directly to the Chevalley
generators ($e_{1},\;f_{1}$). The solutions can be written,
in our notations, as
$$
c_{(\epsilon,\epsilon ')}(j_{2},j_{4},m_{4})=(j_{4}+\epsilon '\;m_{4}+
{1+\epsilon ' \over 2})^{1/2} c_{(\epsilon,\epsilon ')}(j_{2},j_{4})\;\;\;
(\epsilon,\epsilon '=\pm 1) \eqno(13)
$$
with
$$
\openup 2mm
\eqalign{
&c_{(++)}(j_{2},\;j_{4})=c_{(--)}(j_{2}+{1\over 2},\;j_{4}+{1\over 2}) \cr
&=\biggl({(n_{1}+j_{2}+j_{4}+3)(n_{1}-j_{2}-j_{4})(j_{2}+j_{4}+n_{2}+2)
(j_{2}+j_{4}-n_{2}+1) \over (2j_{2}+1)\;(2j_{2}+2)\;(2j_{4}+1)\;
(2j_{4}+2)}\biggl )^{1/2} \cr} \eqno(14)
$$
and
$$
\openup 2mm
\eqalign{
&c_{(+-)}(j_{2},\;j_{4})=-c_{(-+)}(j_{2}+{1\over 2},\;j_{4}-{1\over 2}) \cr
&=\biggl({(n_{1}+j_{2}-j_{4}+2)(n_{1}-j_{2}+j_{4}+1)(j_{2}-j_{4}+n_{2}+1)
(j_{4}-j_{2}+n_{2}) \over (2j_{2}+1)\;(2j_{2}+2)\;(2j_{4})\;
(2j_{4}+1)}\biggl )^{1/2}. \cr} \eqno(15)
$$

But now the $q$-deformation is the problem. As yet solutions have been
obtained for the following two cases.

(i) For $n_{2}=0,\;n_{1}=n$
$$
\openup 2mm
\leqalignno{
&j_{4}=j_{2}=0,\;{1\over 2},\;\cdots,\;{n\over 2} \cr
&c_{(+-)}(j_{2},j_{4},m_{4})=\;c_{(-+)}(j_{2},j_{4},m_{4})=0 \cr}
$$
and
$$
\openup 2mm
\eqalign{
&c_{(++)}(j_{2},\;j_{4},\;m_{4})=([j_{4}+\;m_{4}+1]_{2})^{1/2}\;
c_{(++)}(j_{2},\;j_{4}) \cr
&c_{(--)}(j_{2},\;j_{4},\;m_{4})=([j_{4}-\;m_{4}]_{2})^{1/2}\;
c_{(--)}(j_{2},\;j_{4}) \cr}
$$
where
$$
\openup 2mm
\eqalign{
c_{(++)}(j_{2},\;j_{4})&=c_{(--)}(j_{2}+{1\over 2},\;j_{4}+{1\over 2}) \cr
&=\biggl({[n_{1}+2\;j_{2}+3][n_{1}-2\;j_{2}]
\over [2j_{2}+1]_{2}[2j_{2}+2]_{2}}\biggl )^{1/2} \cr} \eqno(16)
$$
this is straightforward. The factorisation of $m_{4}$-dependance is what one
would expect. One has a {\bf minimal} $q$-deformation (with $q^{2}$-brackets
appearing as well).

(ii) For $n_{1}=n_{2}=n$
$$
j_{2}+j_{4}=n,\;\;j_{2}=0,\;{1\over 2},\cdots,\;n
$$
$$
c_{(++)}(j_{2},\;j_{4},\;m_{4})=\;c_{(--)}(j_{2},\;j_{4},\;m_{4})=0. \eqno(17)
$$

If one tries to impose for $c_{(\pm,\mp)}$ an $m_{4}$-dependance of the type
one expects from the classical expression and the typical minimal deformation
(found for $n_{2}=0$ say) one runs into a contradiction. The following
remarkable solution has been found. One obtains,
$$
\openup 2mm
\eqalign{
&c_{(+-)}(j_{2},\;j_{4},\;m_{4})=([n+1]_{2}-[j_{2}+\;m_{4}+1]_{2})^{1/2}\;
c_{(+-)}(j_{2},\;j_{4}) \cr
&c_{(-+)}(j_{2},\;j_{4},\;m_{4})=([n+1]_{2}-[j_{2}-\;m_{4}]_{2})^{1/2}\;
c_{(-+)}(j_{2},\;j_{4}) \cr} \eqno(18)
$$
where
$$
c_{(+-)}(j_{2})=-c_{(-+)}(j_{2}+{1\over 2})=\biggl({[2j_{2}+1]\;[2j_{2}+2]\over
[2j_{2}+1]_{2}\;[2j_{2}+2]_{2}}\biggl)^{1/2} \eqno(19)
$$
One preserves the correct classical limit. But the $m_{4}$-dependance involves
a strikingly non-minimal $q$-deformation prescription. (This, to our knowledge,
is the first example of this kind.) One can express the square root of the
difference of two brackets (appearing through $m_{4}$-dependance) as a square
root of products of brackets through the identity
$$
[x]_{2}-[y]_{2}=[x-y]{[x+y]_{2}\over [x+y]}.
$$
But now $m_{4}$ appears in the denominator on the right which is again quite
unusual.

{}From the definition of ($e_{4},\;f_{4}$) one now obtains
( with $j_{4}=n-j_{2}$ )
$$
\openup 2mm
\eqalign{
&e_{4}\vert j_{2}\;m_{2}\;j_{4}\;m_{4}>=-(q+q^{-1})
\;\lbrace ([n+1]_{2}-[j_{2}-m_{4}]_{2})\times \cr
&([n+1]_{2}-[j_{2}+m_{4}+1]_{2})\rbrace ^{1/2}
\vert j_{2}\;m_{2}\;j_{4}\;m_{4}+1> \cr
&f_{4}\vert j_{2}\;m_{2}\;j_{4}\;m_{4}>=-(q+q^{-1})
\;\lbrace ([n+1]_{2}-[j_{2}-m_{4}+1]_{2})\times \cr
&([n+1]_{2}-[j_{2}+m_{4}]_{2})
\rbrace ^{1/2}\vert j_{2}\;m_{2}\;j_{4}\;m_{4}-1> \cr} \eqno(20)
$$
(the negative sign arises due to phase conventions.)

For comparison we note that for $n_{2}=0$ ($j_{2}=j_{4}$) one has
$$
\openup 2mm
\eqalign{
&e_{4}\vert j_{2}\;m_{2}\;j_{4}\;m_{4}> \cr
&=(q+q^{-1})\lbrace [j_{4}-m_{4}]_{2}[j_{4}+m_{4}+1]_{2}\rbrace ^{1/2}
\vert j_{2}\;m_{2}\;j_{4}\;m_{4}+1> \cr
&f_{4}\vert j_{2}\;m_{2}\;j_{4}\;m_{4}> \cr
&=(q+q^{-1})\lbrace[j_{4}+m_{4}]_{2}[j_{4}-m_{4}+1]_{2}\rbrace ^{1/2}
\vert j_{2}\;m_{2}\;j_{4}\;m_{4}-1> \cr} \eqno(21)
$$
here the classical limit and the $SU(2)_{q}$ structure associated with
(${e_{4}\over [2]},\;{f_{4}\over [2]}$, $q^{\pm M_{4}}$) are evident. For
$n_{2}=n_{1}$, the commutator $[e_{4},\;f_{4}]$ is more complicated
but, of course, has the same classical limit.

Studying the bases parallely has other interests than providing interesting
exercises in $q$-deformation. We briefly mention two important aspect to be
explored elsewhere.

(a) Suitably adapting familiar continuation techniques $SO(3,2)_{q}$ and
$SO(4,1)_{q}$ representations can be obtained from basis (1) and basis (2)
respectively.

(b) Under suitable contraction procedures again $q$-deformation of
representations of different inhomogeneous algebras are obtained in the two
cases. The contractions of basis (1) are discussed in $[1]$. Contracted
representations arising from basis (2) will be presented elsewhere. Here
possibilities of applications are particularly interesting.

The major remaining task is the explicit construction of $SO(5)_{q}$
representations for arbitrary admissible, ($n_{1}$, $n_{2}$). The elegant
formalism of Fiore $[4]$ gives the deformations of only the vector
representations of $SO(N)$. If one intends to cover the full range of
invariants and indices some essential, hard problems are encounted already
at the level of $SO(5)_{q}$. Overcoming them is the motivation behind our
efforts.

The basis (1) classical representations seem (so far) to permit relatively
simple (minimal) $q$-deformation. But the intricate multiplicity patterns
(presented here for the first time) indicate the difficulties of a ( even
classical) general solution. The unsuitability of the classical Gelfand-Zetlin
representations $[2]$ for $q$-deformation was explained in $[1]$. The
classical representations of Hughes $[3]$ (starting point of our basis (2))
have attractive properties but their $q$-deformation presents unxpected
problems. We hope to present a general solution for basis(2) in a following
paper.

The domains of the indices were considered above for generic $q$. For $q$
a root of unity the situation ( concerning dimensions and the center) changes
radically. Neverthless, the periodic and partially periodic irreducible
representations for $q$ a root of unity can be obtained from generic $q$ ones
using our formalism of {\bf fractional parts }$[5]$ $[1]$. This will not be
discussed here. We refer, however, to section IV of $[1]$ for explanations and
references.

{\bf ACKNOWLEDEGMENT:}

This work is supported in part by the EEC contracts No. SC1-CT92-0792 and
No. CHRX-CT93-0340.

\eject

\centerline{{\bf REFERENCES:}}

\item{1.}
A. Chakrabarti, "$SO(5)_{q}$ and contraction...",
J.Math.Phys. 35, 4247 (1994).

\item{2.}
I.M. Gelfand, R.A. Milnos and Z.Ya. Shapiro, Representations of
Rotation and Lorentz Groups; Pergman, New-york, 1963.(Supplements).

\item{3.}
J.W.B. Hughes: "$SU(2) \otimes SU(2)$ shift operators and
representations of $SO(5)$". J.Math.Phys. 24, 1015 (1983).

\item{4.}
G. Fiore," Realisation $U_{q}(SO(N))$ within the differential algebra
on $R_{q}^{N}$" ( SISSA preprint June 1993).

\item{5.}
The formalism of fractional parts for $q$ a root of unity was first
presented in

\item{}
D. Arnaudon and A. Chakrabarti. Periodic and partially periodic
representations of $SU(N)_{q}$., Comm. Math. Phys. 139, 461 (1991)

\item{}
Ref.1 cites more sources. A different approach (with supplementary references)
can be found in

\item{}
D. Arnaudon and A. Chakrabarti, "Periodic Representations of
$SO(5)_{q}$", Phys.lett. B 262, 68 (1991).

\end